\documentclass[journal,twoside,web]{ieeecolor}
\usepackage{generic}
\usepackage{cite}
\usepackage{amsmath,amssymb,amsfonts}
\usepackage{algorithmic}
\usepackage{graphicx} 
\usepackage{textcomp}
\usepackage{booktabs}
\usepackage{multirow}
\usepackage{color, hyperref}
\usepackage{comment}
\usepackage{enumerate}
\usepackage{hyperref}
\usepackage{bm}
\usepackage{mathtools}
\usepackage[T1]{fontenc}

\def\BibTeX{{\rm B\kern-.05em{\sc i\kern-.025em b}\kern-.08em
    T\kern-.1667em\lower.7ex\hbox{E}\kern-.125emX}}
\begin{document}
\title{Limited-Angle CT Reconstruction Using Multi-Volume Latent Consistency Model}
\author{Hinako Isogai, Naruki Murahashi, Mitsuhiro Nakamura, Megumi Nakao
    \thanks{This work involved human subjects or animals in its research. Approval of all ethical and experimental procedures and protocols was granted by the Kyoto University Certified Review Board (approval number: R1446-3), and performed in line with the Declaration of Helsinki. This research was supported by JSPS Grant-in-Aid for Scientific Research (B) (grant number 19H04484 and 22H03021).}
    \thanks{The authors are with the Human Health Sciences Graduate School of Medicines, Kyoto University. Shogoin, Sakyo, Kyoto, 606-8507, JAPAN.}}
\maketitle

\begin{abstract}
Limited-angle computed tomography (LACT) reconstruction is an inverse problem with severe ill-posedness arising from missing projection angles, and it is difficult to restore high-precision images without sufficient prior knowledge. In recent years, machine learning methods represented by diffusion models have demonstrated high image generation capabilities, and their application to sparse-view CT and LACT reconstruction has progressed. However, accurate restoration of three-dimensional structures of organs and vessels and preservation of contrast remain challenges, and the impact of differences in diverse clinical imaging conditions such as field of view (FOV) and projection angle range on reconstruction accuracy has not been sufficiently investigated.

In this study, we propose a multi-volume latent diffusion model that uses three-dimensional latent representations obtained from multiple effective fields of view as guidance for LACT reconstruction in clinical practical problems. The proposed method achieves fast and stable inference by introducing consistency models into latent space, and enables high-precision preservation of organ boundary information and internal structures under different FOV conditions through a Multi-volume encoder that acquires latent variables from different scales of the global region and central region.

The evaluation experiments demonstrated that the proposed method achieved high-precision synthetic CT image generation compared to existing methods. Under the limited-angle condition of $60^\circ$, mean absolute error (MAE) of 10.12 HU and structual similarity (SSIM) of 0.9677 were achieved, and under the extreme limited-angle condition of $30^\circ$, MAE of 16.69 HU and SSIM of 0.9393 were achieved. Furthermore, stable reconstruction performance was demonstrated even for unknown projection angle conditions not included during training, confirming the applicability to diverse imaging conditions in clinical practice.
\end{abstract}

\begin{keywords}
Latent diffusion model, Limited angle CT reconstruction, Image synthesis.
\end{keywords}

\maketitle

\section{Introduction}
\label{sec:introduction}
X-ray computed tomography (CT) is an imaging modality that reconstructs cross-sectional images from X-ray projections acquired from multiple directions \cite{Kak2001}. Three-dimensional (3D) CT imaging enables visualization of internal anatomical structures with high spatial resolution and quantitative accuracy, and is widely used in clinical practice \cite{Chiffre2014, Wang2018, Wang2008}. However, conventional CT systems are typically designed as fixed installations, which limits their applicability in situations requiring immediacy or portability, such as image-guided radiotherapy or intraoperative guidance. To address these constraints, compact and fast imaging systems such as cone-beam CT (CBCT), as well as on-demand imaging based on two-dimensional X-ray projections, have attracted increasing attention. In addition, limited-angle X-ray imaging, exemplified by tomosynthesis in mammography, has been actively investigated for clinical applications \cite{Helvie2010} and has stimulated further miniaturization of imaging devices. Nevertheless, limited-angle X-ray acquisition inherently lacks sufficient information for accurate reconstruction, and therefore its clinical use is often restricted to qualitative interpretation of two-dimensional images.

Recently, CT reconstruction from incomplete projection data via machine learning techniques has been extensively studied. Two representative problem settings are sparse-view CT reconstruction, which reconstructs images from sparsely sampled projections over a full angular range \cite{sparseChen2017,sparseZheng2017,sparseYe2018,sparseZhang2018}, and limited-angle CT (LACT) reconstruction, which recovers images from projection data acquired only within a restricted angular range \cite{LACTWuerfl2018,LACTChen2022,LACTZhao2018,LACTWang2021}. Sparse-view CT is primarily motivated by dose reduction, where projections are acquired from all directions, but with large angular intervals that provide insufficient sampling density. Under these conditions, classical reconstruction methods such as filtered backprojection (FBP) suffer from severe streak artifacts. In contrast, LACT reconstruction assumes that projections are available only within a limited angular range owing to physical or environmental constraints of the imaging system, such as dental CT or intraoperative CBCT. Conventionally, CBCT reconstruction is based on the Feldkamp–Davis–Kress (FDK) algorithm \cite{FDK1984}, which is often combined with iterative reconstruction techniques \cite{TVSidky2008, AnisotropicTVJin2010}. However, depending on the acquisition conditions, these methods tend to result in strong artifacts and locally missing structures. In particular, LACT imaging with an angular coverage less than $180^\circ$, which is the focus of this study, constitutes a fundamentally ill-posed reconstruction problem owing to the severe lack of directional information unless appropriate prior knowledge is incorporated.

With the rapid progress of deep learning, numerous learning-based approaches for CT reconstruction from incomplete projection data have been reported. These include convolutional neural network (CNN)-based denoising models \cite{Zhang2017DnCNN}, end-to-end image generation networks \cite{Jin2017FBPConvNet}, iterative reconstruction methods \cite{Huang2013NDTV,Hu2017ImprovedSILimitedSparse}, and adversarial learning approaches using generative adversarial networks (GANs) \cite{Goodfellow2014GAN,Wolterink2017GAN,Nakao2020Accessgan,Nakamura2021gan,Hase2021gan} for artifact reduction. More recently, diffusion models \cite{SohlDickstein2015} have demonstrated remarkable image generation capabilities and have been applied to medical image reconstruction\cite{Han2025PhysicsInformedScore}. Diffusion-based approaches for sparse-view CT and LACT reconstruction have been proposed and have achieved high-quality images \cite{Sun2024DIFR3CT,LACTDOLCELiu2023,Yang2024DualDomainDiffusionSVCT}. However, accurate recovery of 3D anatomical structures and the preservation of image contrast remain challenging. Moreover, the effects of clinically relevant acquisition conditions, such as the field of view (FOV), projection direction, and angular range on the generated images have not yet been sufficiently investigated.

Here, we aim to address LACT reconstruction under clinically realistic conditions via a multi-volume latent-diffusion model guided by latent variables extracted from multiple effective fields of view. To overcome the long inference time associated with conventional denoising diffusion probabilistic models (DDPMs), our method adopts a consistency model \cite{LCMLuo2023} that enables fast image generation. Furthermore, by transforming 3D CT images into latent representations and performing the reverse diffusion process in the latent space, efficient high-resolution image generation can be obtained. However, when the FOV is large, reconstructed images can lack fine anatomical details. To address this issue, we developed a multi-region-of-interest encoder that extracts latent variables from image regions at substantially different spatial scales, including both global and local regions. By incorporating additional 3D latent information from slices outside the target slice, the framework aims to produce more accurate reconstructions of 3D anatomical structures.

To evaluate the effectiveness of LACT reconstruction, 3D CT datasets were acquired from 135 patients who underwent radiotherapy for pancreatic cancer at Kyoto University Hospital. Initially, with $60^\circ$ limited-angle imaging, we compared different types of guidance provided to the diffusion model to analyze the characteristics of the multi-volume latent representation. Next, the method was quantitatively and qualitatively compared with conventional deep-learning-based reconstruction approaches to demonstrate its effectiveness. Finally, assuming realistic clinical acquisition scenarios, we investigated generalization under different angular conditions and analyzed reconstruction results under extremely limited angles below $60^\circ$. In this way, we clarified the performance limits of LACT reconstruction.

\begin{figure*}[t]
     \centering
     \includegraphics[width=0.9\linewidth]{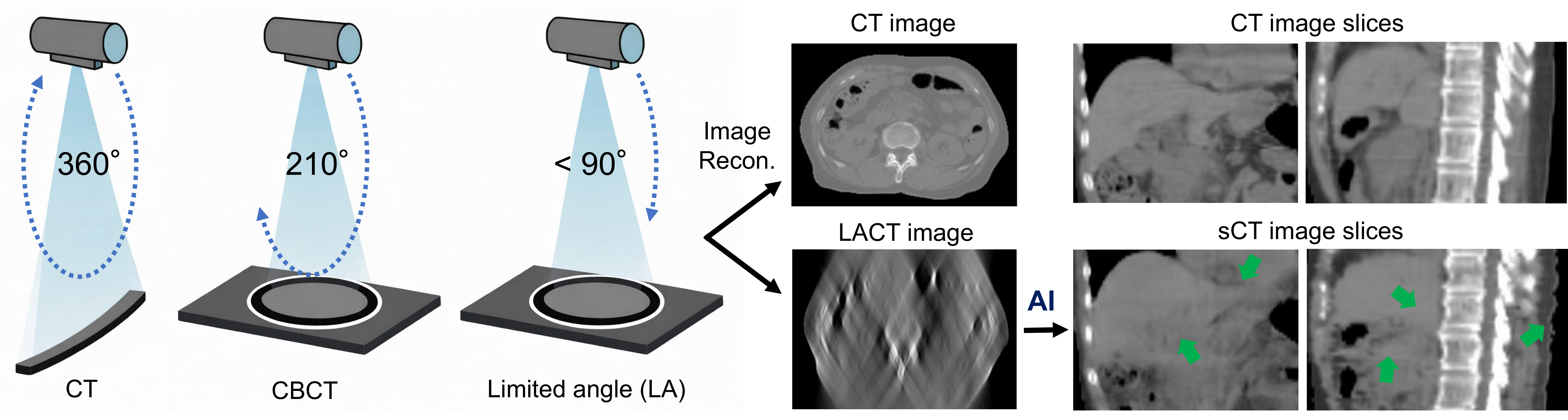}
     \caption{ Computed tomography (CT)/cone-beam CT (CBCT) reconstruction and limited-angle CT (LACT) image generation using generative artificial intelligence (AI). Synthetic CT (sCT) images obtained via generative AI-based LACT image completion remain challenging. Green arrows highlight incorrectly reconstructed voxels exhibiting low contrast and rough surface structures. }
     \label{fig1}
\end{figure*}

The main contributions of this work are summarized as follows:
\begin{itemize}
  \item We developed a latent consistency model for LACT reconstruction, in which 3D latent variables extracted from regions outside the target slice are used as guidance.
  \item By leveraging latent representations at different global and central region spatial scales, the method enabled high-contrast reconstruction of 3D anatomical structures under different FOVs.
  \item We analyzed the image generation performance and the generalization capability under different angular settings, including severely limited angles below $60^\circ$.
\end{itemize}

\section{Related Works}
\label{sec:related_works}

\subsection{CT image reconstruction}

CT image reconstruction has long been studied as an inverse problem of the Radon transform. Analytical reconstruction methods, such as FBP, have been widely used as a standard approach in medical CT, while the FDK algorithm \cite{FDK1984} is commonly adopted for CBCT. However, in LACT or sparse-view CT, where the projection angles or the number of projections are restricted, FBP and FDK often suffer from severe image degradation, including streak artifacts, blurring, and structural distortions.

To mitigate these issues, physics-based iterative reconstruction methods have been extensively developed \cite{Huang2013NDTV,TVSidky2008,Chen2013AnisoTV,AnisotropicTVJin2010,Gong2019RelativeTVLACT,Mohan2015TIMBIR,Venkat2013MBIR_STEM}. They improve image quality by incorporating regularization based on prior assumptions, but practical deployment is often hindered by the challenge of designing appropriate regularizers that can incur high computational costs. As a result, there has been a rapid shift toward data-driven deep-learning-based approaches.

\subsection{Sparse-view CT reconstruction}

In sparse-view CT, the number of projections is severely limited, leading to significant information loss in both the projection and image domains. Consequently, deep-learning-based artifact reduction methods have been investigated. Representative approaches include CNN and U-Net-based frameworks that suppress streak artifacts in reconstructed images \cite{Jin2017FBPConvNet,Han2018FrameletUNet}.

GANs \cite{Goodfellow2014GAN} have also been widely explored in this context. Wolterink et al.~\cite{Wolterink2017GAN} introduced GAN-based denoising for low-dose CT and demonstrated that generative models are effective in restoring realistic textures. Han et al.~\cite{Han2019X2CTGAN} proposed X2CT-GAN that estimates volumetric CT images from only two X-ray projections, demonstrating that high-quality image generation is possible from extremely sparse inputs.

Dual-domain approaches jointly exploit the projection (sinogram) and image domains. Wu et al.~\cite{Wu2021DRONE} proposed DRONE , which combines dual-domain CNNs with data-consistency optimization for error-component learning. This suppresses streak artifacts caused by sparse-view acquisition while enabling the recovery of fine structural details.

More recently, generative modeling approaches have advanced. For example, Guan et al.~\cite{Guan2024SGM} introduced a score-based generative model operating in the sinogram domain that enabled high-accuracy reconstruction via sequential generation. These studies demonstrate that various deep generative approaches have become active directions for sparse-view CT reconstruction.

\subsection{Limited-angle CT reconstruction}

The fundamental difference in reconstruction challenges between sparse-view CT and LACT is the spatial extent of missing information. As illustrated in Figure ~\ref{fig1}, LACT suffers from a large missing wedge due to restricted acquisition angles, making reconstruction substantially more challenging (while sparse-view CT samples projection data over the full angular range). As shown in LACT and synthetic images, conventional reconstruction methods inevitably produce strong directional streak artifacts and structural deficiencies. Even with deep-learning-based approaches, hallucinated structures may appear.

These limitations cannot be fundamentally resolved by analytical methods such as FBP/FDK or by classical iterative reconstruction. Even though they provide stable, physics-based reconstruction frameworks, they exhibit inherent limitations in recovering missing structures in large angular gaps. Therefore, they are insufficient for addressing the core challenges of LACT reconstruction.

Deep-learning-based approaches for artifact reduction and sinogram completion have shown substantial improvements. For example, Hu et al.~\cite{DIOR2022Hu} used a deep reconstruction framework that combined iterative optimization with neural-network-based residual correction to suppress directional streak artifacts by explicitly learning residual components. Pan et al.~\cite{LACTPan2024} used an unrolled framework that alternated between physics-based updates and network-based residual refinement for focusing on error-domain optimization to enable both high data consistency and improved reconstruction accuracy. However, for large missing angular ranges, hallucinated structures remain unavoidable, and morphological fidelity becomes increasingly difficult.

Generative models, particularly diffusion-based approaches, have been used for LACT reconstruction.

Diffusion models \cite{SohlDickstein2015} consist of a forward diffusion process that gradually adds Gaussian noise and reverse denoising that progressively generates samples. Variants such as DDPM  \cite{Ho2020}, denoising diffusion implicit models (DDIM) \cite{Song2020}, and latent diffusion models \cite{Rombach2022} have demonstrated remarkable image generation. A key advantage of diffusion models lies in the ability to recover global structures before refining details via multi-step denoising, as well as their flexibility in incorporating observation-based conditioning. These advantages make diffusion models particularly suitable for medical image reconstruction, where physical models can be embedded into the inference process. Beyond unconditional generation, diffusion models have also been widely applied in various conditional image-to-image translation tasks \cite{Saharia2022Palette,Xia2022LowDoseDDPM,Saharia2022IterativeRefinement}.

Diffusion models are thus considered promising frameworks for addressing severely ill-posed problems such as LACT, where substantial portions of information are missing.

Several studies have applied diffusion models specifically to LACT reconstruction. Liu et al.~\cite{LACTDOLCELiu2023} used DOLCE (Diffusion Probabilistic Limited-Angle CT Reconstruction), which combined transformer-based sinogram completion with an image-domain refinement network in a dual-domain framework. By stabilizing global structure completion in the missing angular range and suppressing artifacts via dual-domain processing, DOLCE enabled state-of-the-art performance for $60^\circ$ angular constraints. Nevertheless, limitations included reduced reconstruction fidelity for anatomically complex abdominal organs, a strong dependence on the trained angular range, and no explicit enforcement of 3D structural consistency.

Wang et al.~\cite{LACTWang2024} introduced time-reversal sampling to accelerate score-based reconstruction. While this approach improved generation efficiency, it mainly focused on computational speed and did not directly address the faithful recovery of high-frequency structures or internal organ morphologies.

Han et al.~\cite{Han2025PhysicsInformedScore} used a physics-informed score-based diffusion model for LACT reconstruction in coronary CT angiography, which integrated diffusion score functions with physics-based iterative optimization. By guiding denoising toward consistency with measured data, their method suppressed streak artifacts in $90^\circ$ and $120^\circ$ angular ranges. However, because the approach did not explicitly complete missing sinogram information, its ability to recover high-frequency structures was limited for large missing angular ranges.

In summary, all of these limitations highlight the need for new approaches that can more faithfully reconstruct high-frequency details and internal anatomical structures under conditions of severely limited angles. This is the primary motivation of the work here.

\section{Methods}
\label{sec:methods}

\subsection{Problem Definition and Preprocessing}
\label{subsec:problem_setting}

We address the problem of generating images with qualities comparable to CT images obtained under full projection conditions through image transformations via diffusion models in limited-angle X-ray imaging. In our mathematical notations, we define the CT image obtained by full-view imaging as the reference image $y_0$, and the LACT image reconstructed from X-ray projection images acquired under LA acquisition as $x$. The objective is to generate a synthetic CT (sCT) image $\hat{y_0}$ with quality close to $y_0$, conditioned on the LACT image $x$.

We formulate the above problem as a self-supervised learning framework.
Specifically, we artificially generate LACT images $x$ by assuming LA imaging from CT images $y_0$ in advance. We perform training using $x$ as a guide for the diffusion model, while using $y_0$ as the teacher signal for restoration.
By using the trained model whose weights are optimized via learning, sCT images $\hat{y_0}$ with features close to CT images can be obtained at the inference time by using only LACT images. However, as shown in Figure \ref{fig1}, LACT reconstruction is a severely ill-posed problem when the angular gaps are large, and simple generative models are prone to generating nonexistent structures (hallucination) and breakdowns of 3D structures of organs and vessels \cite{Sun2024DIFR3CT}.

To address this challenge, we introduce a conditional latent consistency model (CLCM) as the generative model and combine it with multi-volume encoding focused on organ internal structures. This accommodates differences in the FOV during imaging, and enables reconstruction that simultaneously restores high-frequency features and 3D structures.

\begin{figure}[t]
    \centering
    \includegraphics[width=90mm]{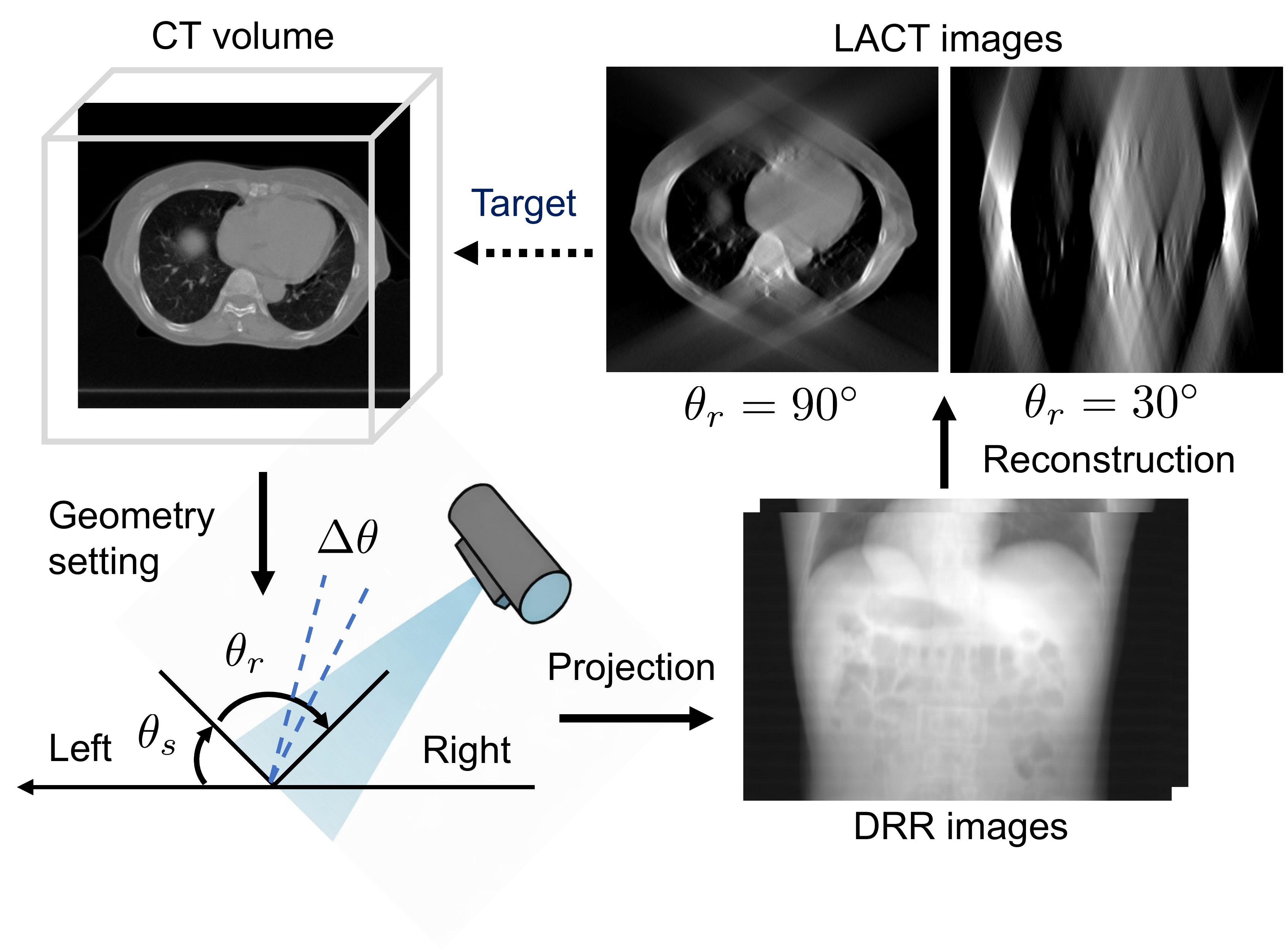}
    \caption{Flow of limited-angle computed tomography (LACT) image simulation for self-supervised learning of image completion. $\theta_s$: angle at projection start, $\theta_r$: angular range from projection start to end and $\Delta \theta$: angular interval between each projection image.}
    \label{fig2}
\end{figure}

Figure \ref{fig2} shows the flow of LACT image simulation used as a guide for the diffusion model. This assumes the case where X-ray imaging similar to CBCT imaging is performed under LA acquisition. In the imaging settings, the left-hand direction in the supine position is defined as $0^\circ$, and imaging under the following angular conditions is performed clockwise:

\begin{itemize}
 \item $\theta_s$: angle at projection start
 \item $\theta_r$: angular range from projection start to end
 \item $\Delta \theta$: angular interval between each projection image
\end{itemize}

\noindent
Here, $\theta_r$ is variable, and we aim to train and validate the inference performance of sCT image restoration tasks for three LA settings targeting wider angle ranges: $\theta_r = 120^\circ, 90^\circ, 60^\circ$, with $\theta_r = 60^\circ$ as the representative setting, which is the minimum angle investigated in \cite{LACTDOLCELiu2023}, and three extreme LA acquisition settings targeting narrower angle ranges: $\theta_r = 45^\circ, 30^\circ, 15^\circ$. Furthermore, to investigate the generalization performance of trained models under LA and extreme LA settings, we also construct LACT image databases for four unseen LA settings used only for testing: $\theta_r = 105^\circ, 75^\circ, 37.5^\circ, 22.5^\circ$. Note that for $\theta_s$ and $\Delta \theta$, considering clinical imaging conditions, we set $\theta_s = -\theta_r / 2 + 90^\circ$ and $\Delta \theta = 0.5^\circ$.

\begin{figure*}[h]
     \centering
     \includegraphics[width=.95\linewidth]{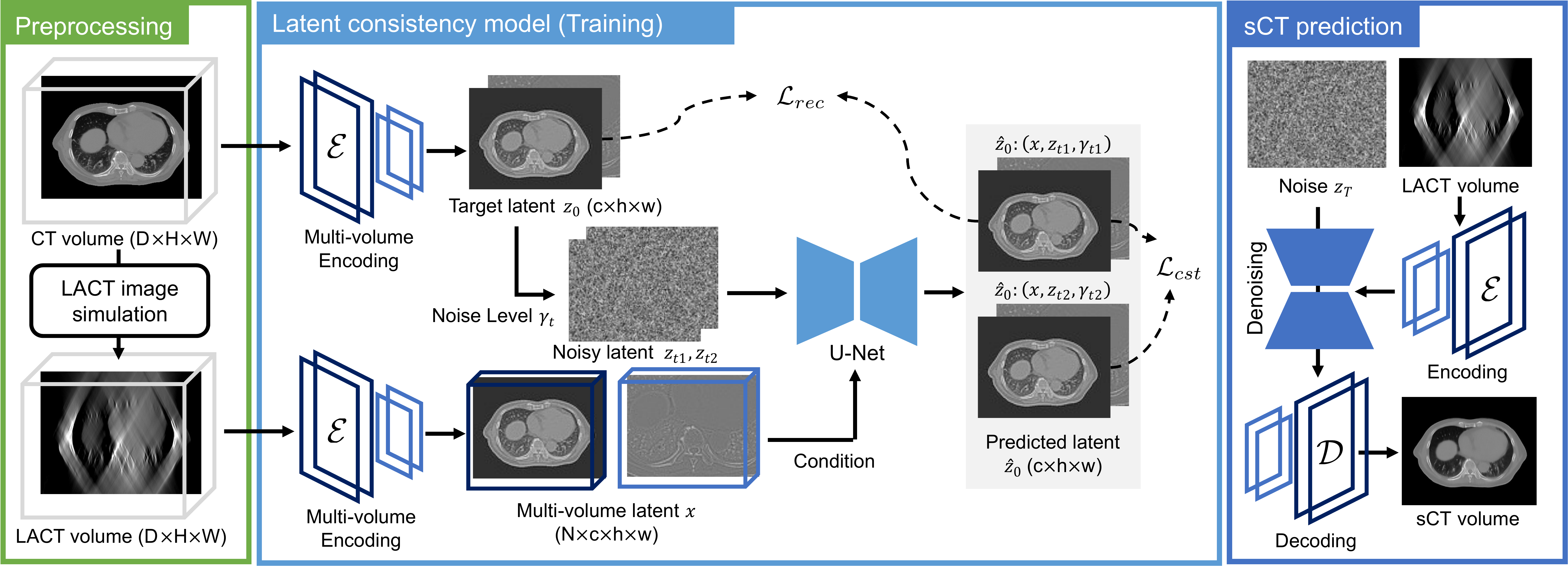}
     \caption{Self-supervised multi-volume conditional latent diffusion model for LACT image restoration. Multi-volume encoding extracts both global anatomical structures and high-frequency components through different convolution pathways from whole and local 3D regions. Restoration from noisy latents at two time points are learned during diffusion training.
performed.} 
     \label{fig3}
\end{figure*}

Using the ten LA settings described above, we generated digitally reconstructed radiographs from CT volumes and obtained LACT volumes from the resulting radiograph image sets via backprojection calculations using the FDK method \cite{FDK1984}. For this series of forward projection and backprojection calculations, we used the TIGRE tomographic reconstruction toolbox \cite{Biguri2016Tigre}. Figure \ref{fig2} shows examples of reconstructed LACT images corresponding to original CT slices with settings of $\theta_r = 90^\circ$ and $30^\circ$. As the angular gap increases under LA conditions, the information necessary for backprojection becomes increasingly insufficient, resulting in reconstructed images with inaccurate CT values and severe structural defects. Streak artifacts dependent on projection direction occur extensively, indicating that information restoration is not possible without prior knowledge. In the transformation from incomplete LACT images to CT images using prior knowledge, the central challenge addressed by this study is the design of guidance that promotes learning to capture correlations related to image features and 3D structures contained in both modalities.

\subsection{LACT Diffusion Model with Multi-Volume Guidance}
Figure \ref{fig3} shows the flow of LACT reconstruction using the proposed CLCM. Conditional diffusion models \cite{Saharia2022Palette} can be utilized as image transformation models by providing the image to be transformed as a guide to the diffusion model. However, because they require numerous sequential denoising steps, the length of inference time poses a challenge for clinical use. Moreover, in highly uncertain generation problems, such as LACT reconstruction, denoising can become unstable.

CLCM is an image generation model that has a conditional latent diffusion model structure using latent representations obtained from LACT images as guidance while completing the reverse diffusion computation in a single step based on consistency models \cite{LCMLuo2023}. Training the model in a dimensionally compressed latent space improves computational efficiency while enabling both organ structure preservation and stability in image generation. The transformation to latent variables is realized via multi-volume encoding, a unique feature of CLCM. This encoder enables simultaneous extraction of both global anatomical structures and high-frequency components contained in CT images via independent convolutions targeting very different image regions covering both whole and local areas within the image. Moreover, by obtaining latent variables three-dimensionally from regions outside the slice to be restored, it enables accurate restoration of 3D organ structures. This encoder is realized by extending the image compression model Vector Quantized Variational Auto Encoder (VQVAE) \cite{Oord2017vqvae,Isogai2025} to multi-view, multi-slice inputs.

In the framework of the LACT diffusion model, LACT images and target CT images are converted to latent variables by a pre-trained multi-volume encoder. They. are then concatenated, and learning is performed using the consistency model in latent space. Specifically, by applying multi-volume VQVAE pre-trained on approximately 27,438 CT slices to the reference CT image $y_0$, we obtain a latent representation $z_0$ composed of two channels (global and local) with a spatial resolution of $128 \times 128$.

During the reverse diffusion learning, we generate noisy latents $z_t$ by adding Gaussian noise corresponding to a random time $t$ to the latent variable $z_0$ and optimize the model weights to restore the original $z_0$ from $z_t$. By providing the multi-volume latent $x$ obtained from the CT volume as guidance, we guide the restoration of sCT images consistent with LACT images. Because CLCM uses consistency models, noisy latents at two time points are actually generated and restoration is performed.

At inference time, a tensor combining random noise and multi-volume latent data obtained from LACT images is input to the consistency model, and by performing one inference step, the latent representation of the sCT image is generated. By restoring the obtained latent representation with the trained multi-volume decoder, the final sCT image is obtained.

\subsection{Multi-Volume Latent Representation}
This section describes the network structure of multi-volume VQVAE  designed to obtain latent representations of 3D CT images. In conventional VQVAE \cite{Oord2017vqvae}, convolution operations are uniformly applied to the entire image region and compressed into low-dimensional, multi-channel latent representations. In LACT reconstruction, the restoration of fine textures and organ vascular structures is important. However, the features of these high-frequency components are easily lost during the transformation to latent representations, resulting in reconstructed images with reduced contrast. To address this problem, multi-volume encoding extracts image regions from views of both global and local regions within the image and converts them into latent variables with two different levels of resolution.

\begin{figure}[t]
    \centering
    \includegraphics[width=0.9\linewidth]{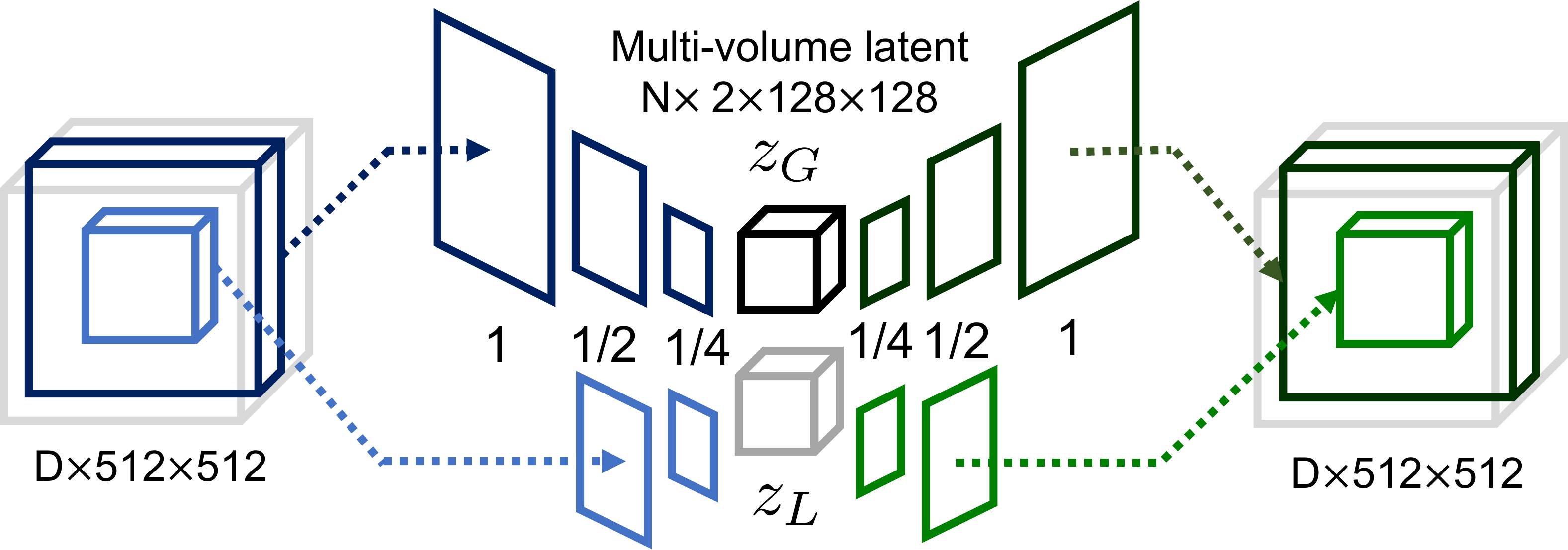}
    \caption{Conceptual overview of the proposed Multi-Volume Vector Quantized Variational Auto Encoder architecture. The two latent representations ${z}_{G}$ and ${z}_{L}$ have the same spatial resolution but encode image features at different resolution levels in the image space.}
    \label{fig4}
\end{figure}

Figure \ref{fig4} shows the multi-volume encoder/decoder structure. It has different convolution pathways for the entire image region and the central region. Initially, from a 3D CT image of $D\times 512\times512$ voxels, we set as the encoding target a sub-volume ${x_G}\in\mathbb{R}^{N\times512\times512}$ composed of a total of $N$ slices with the slice to be restored as the middle slice. Additionally, we let ${x}_{L}\in\mathbb{R}^{N\times256\times256}$ be the local volume extracted from the central region of ${x_G}$.

Next, ${x_G}$ is input to the global path aimed at extracting global structures, and ${x}_{L}$ is input to the local path aimed at extracting local structures. In the global path, convolution and downsampling are applied twice, reducing the spatial resolution by two stages. Feature representations are refined through Residual Stack software to obtain the latent feature representation of global structures ${z}_{G}\in\mathbb{R}^{N\times128\times128}$. In the local path, convolution and downsampling are applied only once to ${x}_{L}$. In this way, feature transformation is performed while preserving high-frequency regions such as internal organ structures, yielding a feature representation with the same spatial resolution as ${z}_{G}$. After similarly applying Residual Stack, we obtain the latent representation of local features ${z}_{L}\in\mathbb{R}^{N\times128\times128}$.

The two obtained latent representations ${z}_{G}$ and ${z}_{L}$ have the same spatial resolution as latent representations but encode image features at different resolution levels in the original image space. ${z}_{G}$ and ${z}_{L}$ are concatenated in the channel direction and utilized as guidance to the final latent representation ${z}\in\mathbb{R}^{N\times2\times128\times128}$. From considerations of memory capacity and computational cost, we set the channels of ${z}_{G}$ and ${z}_{L}$ to 1, but they may also be converted to latent variables using multiple channels. As a result, $z$ becomes a latent variable representation that holds in parallel structural features at different levels within the 3D CT image while preserving the resolution of the original image as much as possible.

In the decoder, the quantized latent representation is separated into global and local components, and decoding is performed by different networks in the global and local paths. The global component is restored to $256\times256$ using deconvolution layers. The local component is decoded as feature values at the same resolution and is then placed in the central region of the image using zero padding. After concatenation in the channel direction, followed by convolution processing, an output image of $512\times512$ is finally generated.

Through this configuration, a latent representation is realized that compresses global features of the entire image while emphasizing the preservation and restoration of local high-frequency organ structures.

\subsection{Conditional Latent Consistency Model}
\label{subsec:CLCM}
This section describes the diffusion model utilizing multi-volume latent representation and its computational procedure in learning and inference for sCT reconstruction. Consistency models \cite{LCMLuo2023} are a type of diffusion model characterized by learning a function that directly maps from a state at an arbitrary noise level to the original data $x_{0}$. During training, consistency constraints are imposed such that outputs at different noise levels match; by learning to obtain the same result regardless of time, fast image generation in a few steps is enabled. Although the accuracy of generated images is considered inferior, there are no known cases of performance evaluation in LACT reconstruction.

Let $z_{0}$ and $x$ be the latent representations obtained through encoding using the trained multi-volume VQVAE \cite{Isogai2025} from CT images and LACT images, respectively. During training, noise corresponding to two random time points $t=t_{n}, t_{m}$ $(1 \leq t_{n}, t_{m} \leq T-1)$, ($t_{n} \neq t_{m}$) is added to $z_{0}$. Here, $T$ is the time at which the image becomes completely noisy, and we used $T = 1000$ referring to prior work \cite{Murahashi2024}. The model is trained to directly predict $z_{0}$ under the condition $x$ for the latent representations at two time points with added noise. $z_{t}$ is expressed in Equation (\ref{eq:zt1}):

\begin{equation}
   z_{t} = \sqrt{\gamma_{t}}z_{0} + \sqrt{1-\gamma_{t}}\epsilon
    \label{eq:zt1},
\end{equation}
where $\epsilon$ is Gaussian noise and $\gamma_{t}$ is the noise level at time $t$, defined by Equation (\ref{eq:gamma1}):

\begin{equation}
   \gamma_{t} = \prod_{i=1}^{t} (1 - \beta_i), \quad \epsilon \sim \mathcal{N}(0, 1),
    \label{eq:gamma1}
\end{equation}
$\beta_{i}$ serves as a noise scheduler (cosine scheduler\cite{A2021cosine}) that controls the strength of added noise. The loss function is defined by Equation (\ref{eq:loss_tsec:otal}):

\begin{eqnarray}
\label{eq:loss_tsec:otal}
\mathcal{L} &=& \mathcal{L}_{rec} + \lambda_{\mathrm{cst}} \mathcal{L}_{cst}\nonumber\\ 
&=&\mathbb{E}\Big[
\| \hat{z}_0^{(t_{n})} - z_0 \|^2 
+ \lambda_{\mathrm{cst}} \, 
\| \hat{z}_0^{(t_{n})} - \hat{z}_0^{(t_{m})} \|^2\Big].
\end{eqnarray}

Thus, the weights are updated during training to minimize the sum of the squared error between $\hat{z}_0^{(t_{n})}$ predicted from the latent representation $z_{t_{n}}$ at time $t=t_{n}$ and $z_{0}$ at the initial time, and the squared error between $\hat{z}_0^{(t_{n})}$ and $\hat{z}_0^{(t_{m})}$ predicted from times $t=t_{n}, t_{m}$. Here, $\lambda_{\mathrm{cst}}$ is the coefficient of the consistency loss, and we used $0.1$ considering the balance with the reconstruction loss. This enables predictions of consistent latent representations across multiple time points. This reduces numerous iterative inference steps required in conventional diffusion models, and enables fast latent representation restoration with fewer steps.

At inference times, latent representations are first obtained from LACT images using the trained multi-volume VQVAE. Next, denoising is performed in one step on internally generated Gaussian noise using the obtained latent representation of the LACT image as guidance. Finally, that latent representation is decoded by the trained multi-volume VQVAE to obtain the final sCT image. The above inference processing is performed on a slice-by-slice basis, but the same initial noise is used for all slices to enable three-dimensionally continuous image completion.

\section{Experiments}
\label{sec:experiments}

To evaluate the model, we conducted experiments to verify: 1) the impact of multi-volume guidance on LACT reconstruction, 2) an inter-model comparison of reconstruction performance, and 3) the generalization performance for unseen projection angle intervals. The model was implemented using Python 3.12.10 and PyTorch 2.8 software, and all experiments were conducted on a computer configured with an Intel Core Ultra 9 285-K 3.70-GHz CPU with 64.0-GB of memory, and a NVIDIA GeForce 5090 32-GB GPU.

\subsection{Dataset and Experimental Settings}
The dataset used for model training and inference used 3D CT volume data of the abdominal regions of 135 patients receiving radiation therapy for pancreatic cancer at Kyoto University Hospital. This was approved by the Ethics Committee of Kyoto University Graduate School of Medicine (Approval Number: R1446-3).  The 3D CT images were 512×512-pixel arrays with 93–174 slices. Each slice had a resolution of $1.07\,\mathrm{mm} \times 1.07\,\mathrm{mm}$ and a slice interval of $2.5\,\mathrm{mm}$. In all experiments, values in the CT images were clipped to the range $[-1000\,\mathrm{HU}, 1000\,\mathrm{HU}]$, where HU denotes Hounsfield units, and then normalized to $[-1,1]$. The multi-volume encoder for conversion between CT images and latent variables used a model that was pre-trained on a database of 27,438 CT slices of the abdominal regions from previous work \cite{Isogai2025}.

The abdominal 3D CT database was divided into 118 cases for training data, four cases for validation data, and thirteen cases for test data, and the model was trained using the Adam optimizer with eight or sixteen batch sizes and a learning rate of $1.0\times10^{-4}$. Because a convergence of loss values was observed after approximately 300 epochs under all training conditions, the maximum epoch was set to 400, using the model that achieved the best mean absolute error (MAE) on the validation data. The MAE is the average absolute value of pixel differences between generated and reference images across all slices, and is defined in Equation (\ref{eq:mae}) for two images $X, Y$ with size $N$:

\begin{equation}
\mathrm{MAE}
=
\frac{1}{N^2}
\sum_{i=1}^{N}
\sum_{j=1}^{N}
\left|
X(i,j) - Y(i,j)
\right|.
\label{eq:mae}
\end{equation}
We use structural similarity (SSIM) as an evaluation metric during testing. It evaluates similarities in small regions in two images, such as brightness, contrast, and structure. SSIM is defined by Equation (\ref{eq:ssim}):
\begin{equation}
\mathrm{SSIM}(X,Y)
=
\frac{(2\mu_X \mu_Y + C_1)(2\sigma_{XY} + C_2)}
{(\mu_X^2 + \mu_Y^2 + C_1)(\sigma_X^2 + \sigma_Y^2 + C_2)}
\label{eq:ssim}
\end{equation}
where $X(i,j)$ and $Y(i,j)$ represent pixel values at position $(i,j)$ in images $X$ and $Y$. $N$ is the number of pixels on one side of the image, $\mu_X, \mu_Y$ are the mean pixel values, $\sigma_X, \sigma_Y$ are the standard deviations, and $\sigma_{XY}$ represents the covariance. Values of SSIM closest to 1.0 indicate the highest structural similarity between the reference image and the reconstructed image. In this experiment, we used the scikit-image structural similarity function for the SSIM calculation, using default values for constants and window size.

In addition to evaluating the entire $512\times512$ image, we calculated evaluation metrics for the central $256\times256$ pixels. This region of interest (ROI) was used to more rigorously evaluate cases where the imaging FOV is wide and structures appear only in the central area of the image.

\subsection{Impact of Multi-Volume Guidance}
We verified the effectiveness of multi-volume guidance using LA images obtained with $\theta_r = 60^\circ$. The model is characterized by multi-view encoding, which obtains latent variables from multiple views, and multi-slice encoding, which obtains latent variables from slices other than the target slice for reconstruction. To investigate the performance of each, we prepared trained CLCMs by only modifying the encoder and decoder configurations, and then evaluated the resulting CT images.

\subsubsection{Analysis of Multi-View Encoding}
To investigate the impact of multi-view encoding on generation accuracy, we generated sCT images with guidance fixed with only latent variables obtained from one LACT slice corresponding to the target reconstruction. We conducted a comparative experiment with encoders and decoders configured with two types: conventional VQVAE (single-view) and multi-volume VQVAE (multi-view). All other training conditions were kept the same.

\begin{table}[t]
    \caption{Impact of multi-view encoding on synthetic computed tomography image quality.}
    \centering
    \begin{tabular}{c cc cc}
        \hline
        \multirow{2}{*}{Encoding}
        & \multicolumn{2}{c}{$512\times512$}
        & \multicolumn{2}{c}{$256\times256$ (ROI)} \\
        \cline{2-5}
        & MAE [HU] $\downarrow$ & SSIM $\uparrow$
        & MAE [HU] $\downarrow$ & SSIM $\uparrow$ \\
        \hline
        Single-view
        & 12.26 & 0.9586 & 44.07 & 0.8385 \\
        Multi-view
        & \textbf{11.11}& \textbf{0.9637} & \textbf{39.97} & \textbf{0.8591} \\
        \hline
    \end{tabular}
    \label{tab:multivqvae}
\vspace{0.5cm}    
    \caption{Impact of multi-slice encoding on synthetic CT image quality.}
    \centering
    \begin{tabular}{c cc cc}
        \hline
        \multirow{2}{*}{$N$} 
        & \multicolumn{2}{c}{$512\times512$} 
        & \multicolumn{2}{c}{$256\times256$ (ROI)} \\
        \cline{2-5}
        & MAE [HU] $\downarrow$ & SSIM $\uparrow$  
        & MAE [HU] $\downarrow$ & SSIM $\uparrow$  \\
        \hline
        1  & 11.11 & 0.9637 & 39.97 & 0.8591 \\
        11 & 10.34 & 0.9672 & 36.85 & 0.8723 \\
        31 & \textbf{10.29} & \textbf{0.9672} & \textbf{36.76} & \textbf{0.8731} \\
        61 & 10.36 & 0.9671 & 37.05 & 0.8724 \\
        \hline
    \end{tabular}
    \label{table:guide_number}

\end{table}

Table \ref{tab:multivqvae} shows the median errors between generated and reference images for each VQVAE configuration. We confirmed that MAE and SSIM improved with multi-view encoding when compared with single-view encoding for both the entire image and the central ROI. With multi-view, the MAE was 41.5 [HU]  and SSIM was 0.856, yielding a 5.90\% improvement in MAE.

\begin{figure}[]
     \centering
     \includegraphics[width=0.9\linewidth]{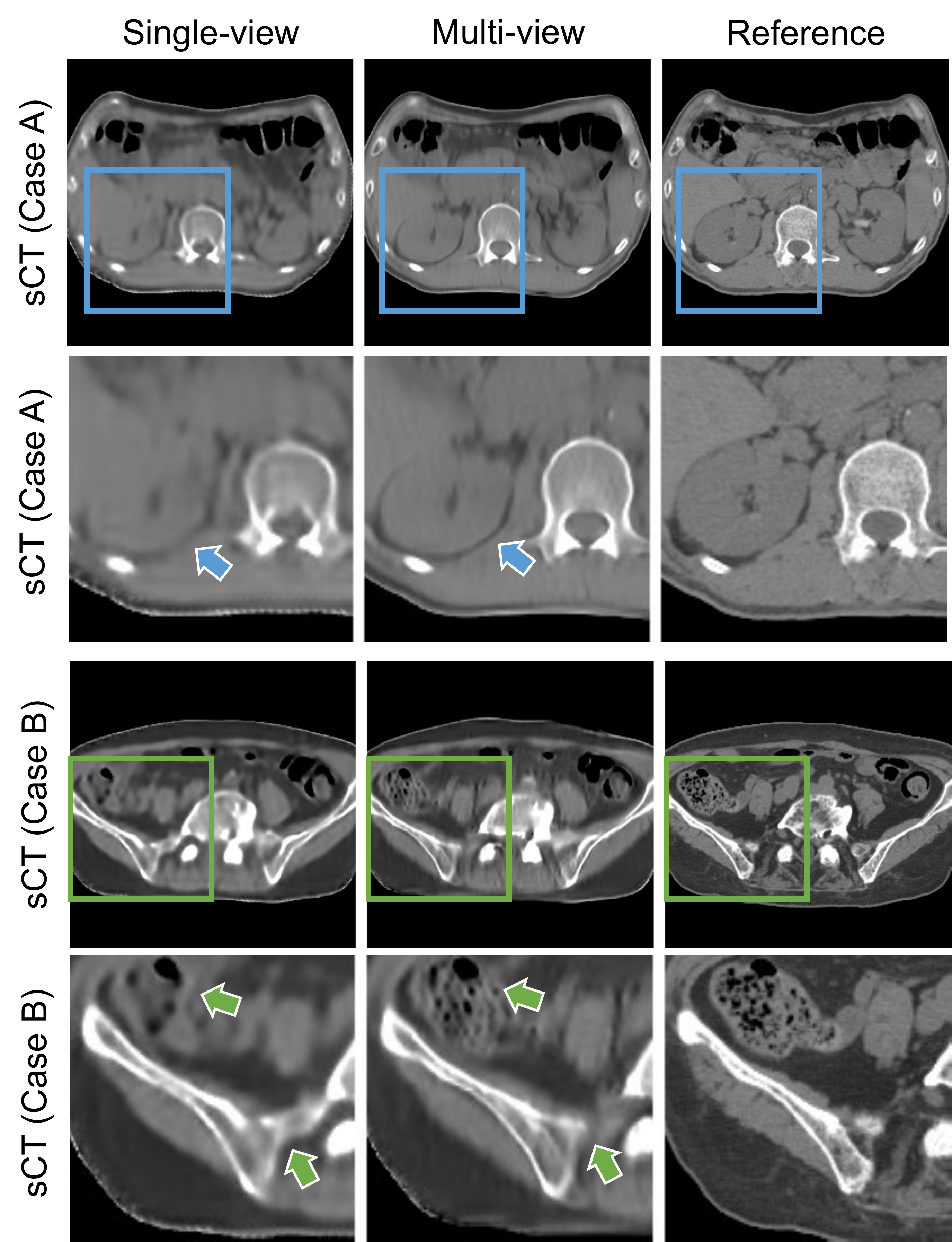}
     \caption{Impact of multi-view encoding on generated images. Impact of multi-view encoding on generated images. Bounding boxes and arrows highlight the improved visibility of organ and skeletal contours achieved with multi-view encoding.}
     \label{fig5}
\end{figure}

Figure \ref{fig5} shows axial sCT images generated by the model and reference images (Reference) for two randomly selected cases (Cases A and B) from the test data. In the abdominal region of Case A around the kidneys, the boundary between the kidneys and surrounding organs is more clearly reproduced with multi-view compared with single-view, thus confirming improved visibility of organ contours. In the pelvic region (Case B), the contours of skeletal structures are expressed more clearly with multi-view. Additionally, the soft-tissue texture in the upper part of the image is unclear with single-view, but fine structures are visible with multi-view. Therefore, multi-view encoding appropriately embeds both overall shape information and local high-frequency structures into the latent space, contributing to improved reproducibility of internal organ structures and boundary information under limited-angle conditions.

\subsubsection{Analysis of Multi-Slice Encoding}
Next, we investigated the impact of the multi-slice encoding feature on image reconstruction accuracy. It acquires latent variables from the superior and inferior directions of the target generation slice and provides the model with 3D structural information of organs as guidance that cannot be obtained from the two-dimensional target reconstruction slice alone. We examined whether this could improve spatial consistency in sCT images.

We used multi-view encoding for the encoder, trained the CLCM by varying the number of slices $N$ provided as guidance, and obtained sCT images. The compared guidance numbers were $N = 1, 11, 31$, and $61$. For example, when $N = 11$, a sub-volume with five additional slices above and below the target generation slice are provided to the diffusion model as guidance.

Table \ref{table:guide_number} shows the median errors between generated and reference images for each slice number. For $N = 31$, MAE was 36.76 [HU] and SSIM was 0.8731, yielding sCT images with the best quality compared with other guidance numbers across all evaluation metrics. No significant difference in quantitative metrics was observed between $N = 31$ and $N = 61$. However, the training time per epoch was approximately 16 min for $N = 31$ and approximately 22 min for $N = 61$. Hence, $N = 31$ was reasonable for both image quality and computational time.

\begin{figure}[t]
    \centering
    \includegraphics[width=0.9\linewidth]{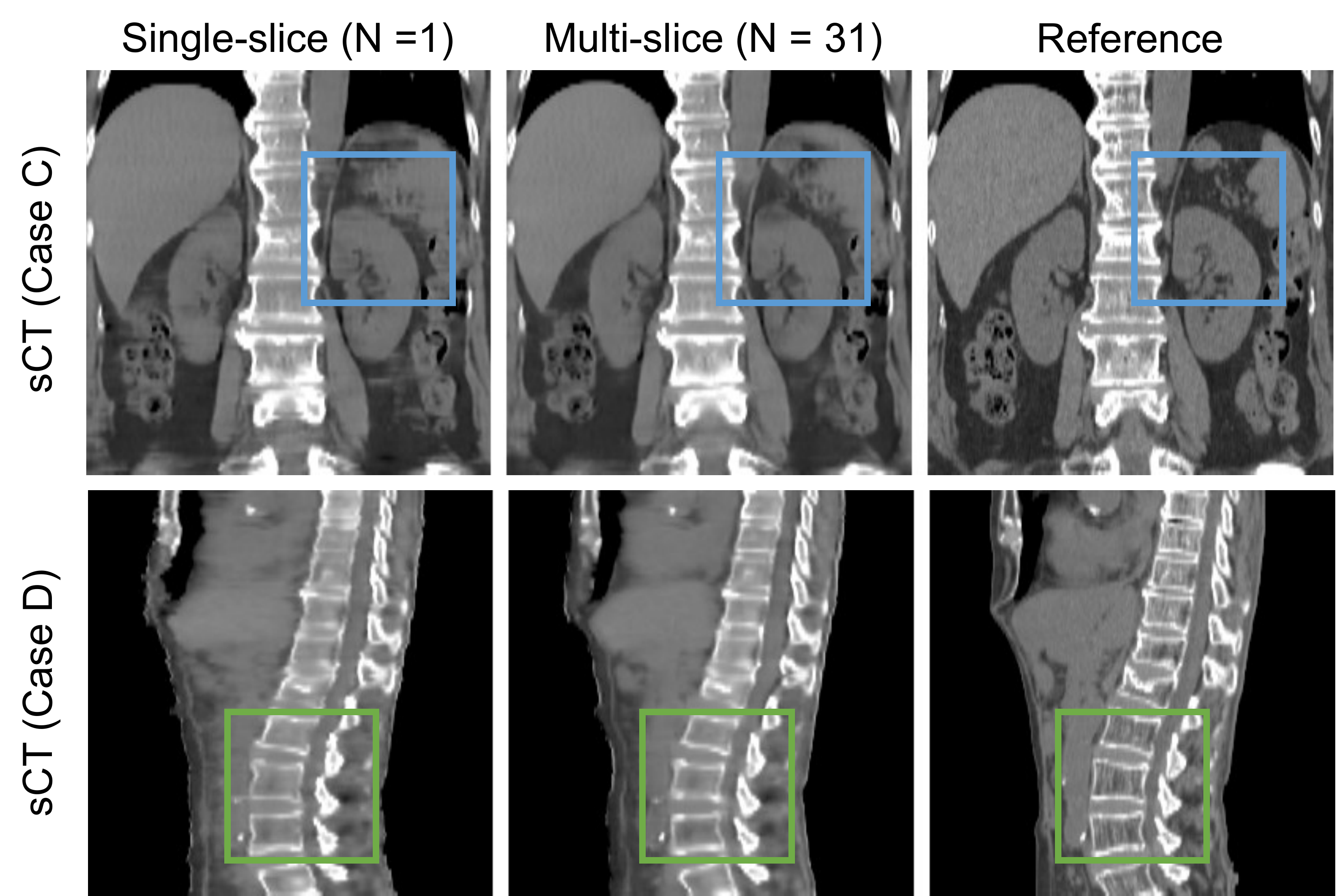}
    \caption{Impact of multi-slice encoding on generated images. Bounding boxes in the coronal and sagittal views demonstrate improved spatial continuity of three-dimensional organ structures.}
    \label{fig6}
\end{figure}

Figure \ref{fig6} shows generated sCT images, reference images (Reference) for two randomly selected cases (Cases C and D) from the test data. 
Case C is an example in the coronal plane. With $N = 1$, locations where 3D organ structure restoration failed were confirmed, as indicated by the blue ROI. In contrast, with $N = 31$, these inconsistencies were reduced and the contours of vertebrae were also reproduced more clearly. 
Case D is an example in the sagittal plane. Compared with $N = 1$, the contours of the vertebrae with $N = 31$ results are depicted more clearly, as indicated by the green ROI. The boundary of the abdominal body surface also appears smoother. Hence, we can confirm image generation with improved reproducibility of 3D organ structures while satisfying spatial continuity of voxels by using multi-slice encoding.

\subsection{Performance Comparison of Models}
\label{sec:comparison}

In this section, we designate the CLCM with multi-view and $N = 31$ multi-volume encoding, which yielded the best results in Experiment 1, as the proposed model and conducted comparative experiments with other existing models. We used pix2pix \cite{isola2017pix2pix}, DDPM \cite{Xia2022LowDoseDDPM}, LDM \cite{Rombach2022}, and LCM \cite{LCMLuo2023} as comparison targets. We implemented training for each model to learn the task of generating sCT images from LACT images on a slice-by-slice basis.

For pix2pix and DDPM, the image size was 256, considering GPU memory capacity and computational time required for training, and the generated images were upsampled to 512 using tri-linear interpolation. 
Note that the spatial resolution of 256 exceeds the latent dimensionality of $2\times128\times128$ adopted in conventional latent diffusion models.
This  provided more advantageous conditions for image generation tasks. Other training conditions were kept as identical as possible to the proposed model, and default values were used for pix2pix hyperparameters.

\begin{table}[t]
    \caption{Median mean absolute error (MAE) and structural similarity index (SSIM) between generated and reference images, compared with existing models. }
    \centering
    \begin{tabular}{c cc cc}
        \hline
        \multirow{2}{*}{Method} 
        & \multicolumn{2}{c}{$512\times512$} 
        & \multicolumn{2}{c}{$256\times256$ (ROI)} \\
        \cline{2-5}
        & MAE [HU] $\downarrow$ & SSIM $\uparrow$  
        & MAE [HU] $\downarrow$ & SSIM $\uparrow$  \\
        \hline
        pix2pix & 10.56 & 0.9373 & 40.76 & 0.7538\\
        DDPM & 33.14 & 0.5223 & 45.66 & 0.6761  \\
        LDM & 13.04 & 0.9517 &  47.95 & 0.8074  \\
        LCM & 12.26 & 0.9586 &  44.07 & 0.8385 \\
        Proposed & \textbf{10.29} & \textbf{0.9672} & \textbf{36.76} & \textbf{0.8731} \\
        \hline
    \end{tabular}
    \label{table:comparison}
\end{table}

Table \ref{table:comparison} shows the median errors and structural similarity between generated and reference images for each method. In both the entire image and the central ROI, the proposed method shows favorable SSIM values compared with the other conventional models, confirming that high-quality sCT image generation is possible.

\begin{figure*}[h]
     \centering
     \includegraphics[width=0.95\linewidth]{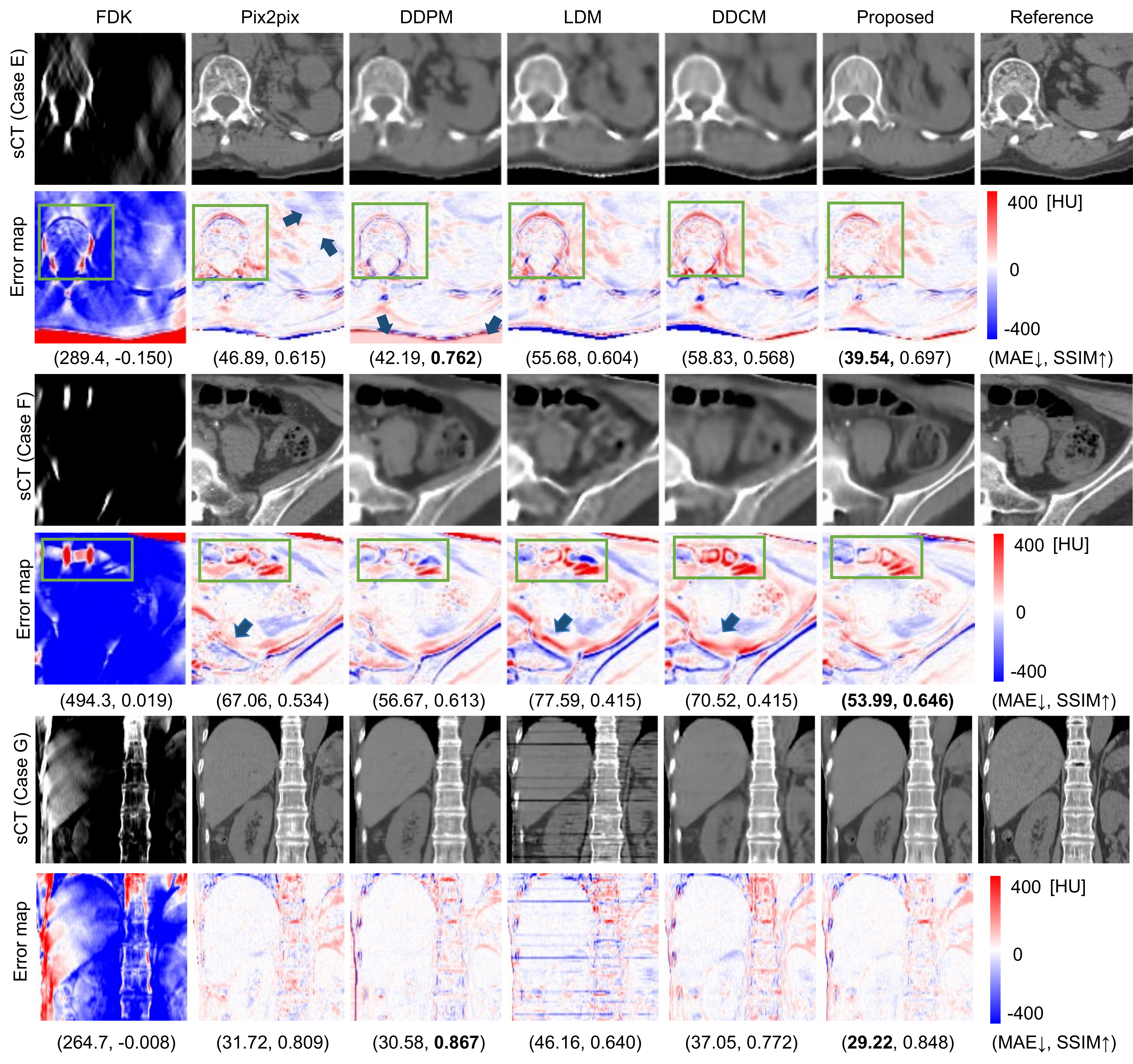}
     \caption{Comparison of synthetic CT ROI images and corresponding subtraction maps across different models. Case E focuses on the kidney region, Case F includes abdominal digestive organs and intestinal gas, and Case G encompasses the abdominal and vertebral regions.}
     \label{fig7}
\end{figure*}

Figure \ref{fig7} shows generated sCT images (Proposed), reference images, and their error map images (Error map) for three randomly selected cases (Cases E, F, and G) from the test data using each method. 
Case E is an axial slice targeting the kidney region. With pix2pix, structures that do not exist in the reference image are generated in the kidney region, indicating hallucination. With DDPM, as shown in the error map image, CT value deviations in the air region persist. For the rectangular ROI in the vertebral region, the contours of vertebral structures are unclear with LDM and LCM. In contrast, the proposed method suppresses these problems, and structural consistency that was difficult to reproduce with other methods is improved. However, regions where kidney contours remain partially unclear persist, indicating that challenges remain in the reproducibility of low-contrast abdominal organs.
Case F is an axial slice targeting the region containing abdominal digestive organs and intestinal gas. With DDPM, intestinal structures in the rectangular ROI between gas regions have disappeared, and the gas regions themselves disappeared with LDM. In contrast, the proposed method generates structures close to the intestinal gas shapes in the reference. 
Case G is a coronal slice including the abdominal and vertebral regions. With pix2pix, artifacts similar to shot noise occur throughout the image, and, with DDPM and LDM, stripe-like artifacts also occur. With DDPM, LDM, and LCM, some inconsistencies are also observed in the 3D structures of organs. In contrast, the proposed method suppresses these artifacts, confirming that the smooth 3D organ structures are restored.

\subsection{Generalization to Unseen Angle Ranges}
\label{sec:angle_range}

\begin{figure}[t]
    \centering
    \includegraphics[width=0.95\linewidth]{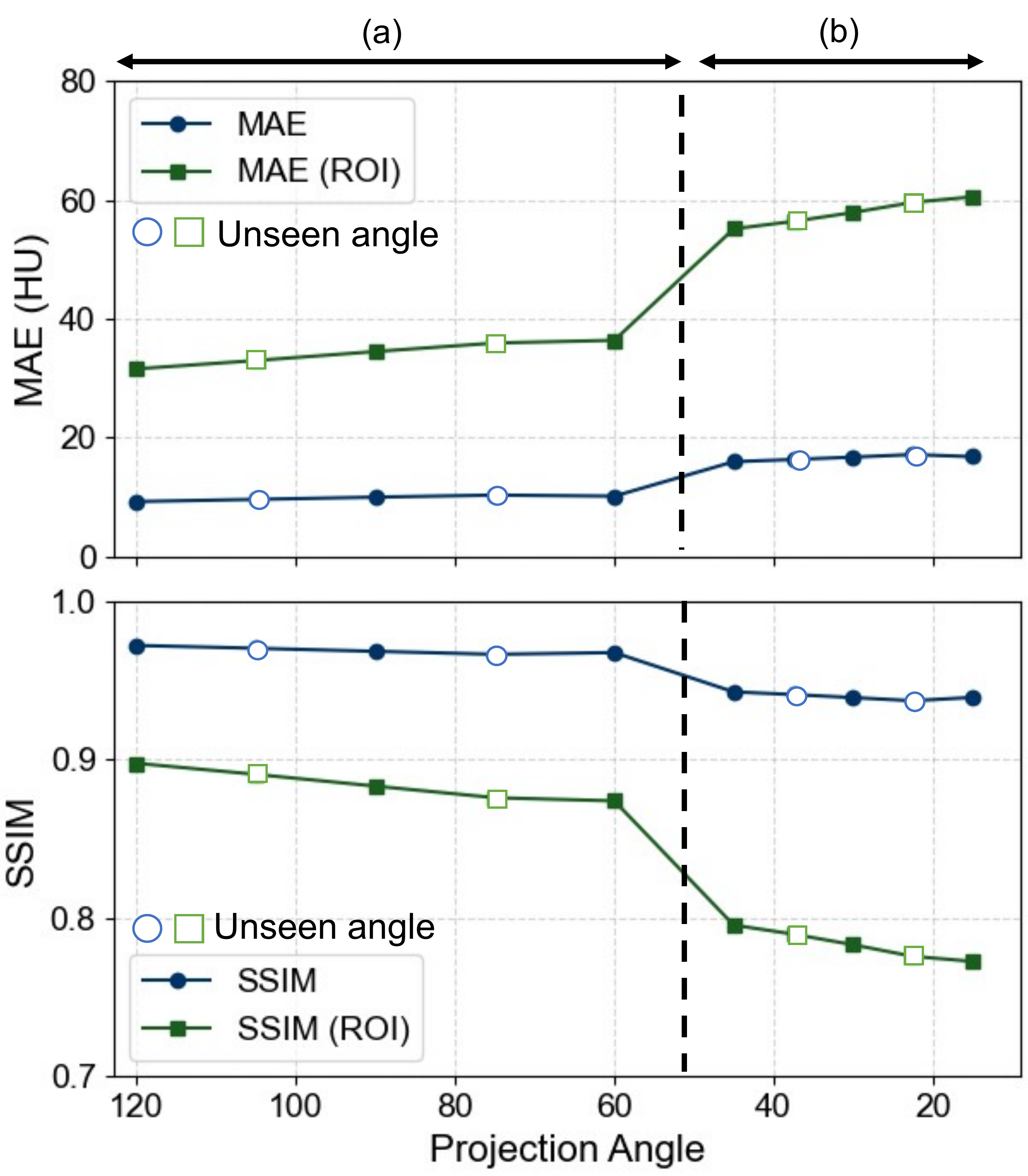}
    \caption{Reconstruction performance of sCT images from LACT images acquired with different projection angle ranges. (a) LA setting and (b) extreme LA setting. The projection angle ranges $\theta_r = 105^\circ, 75^\circ, 37.5^\circ,$ and $22.5^\circ$ correspond to unseen angle ranges.}
    \label{fig8}
\end{figure}

\begin{table*}[t]
    \centering
    \caption{Quality of synthetic CT images for each projection angle in the LA and extreme LA settins. }
    \begin{tabular}{lccccc c ccccc}
        \hline
        & \multicolumn{5}{c}{LA} & & \multicolumn{5}{c}{Extreme LA} \\
        \cline{2-6} \cline{8-12}
        $\theta_r$
        & $120^\circ$ & $105^\circ$ & $90^\circ$ & $75^\circ$ & $60^\circ$
        &
        & $45^\circ$ & $37.5^\circ$ & $30^\circ$ & $22.5^\circ$ & $15^\circ$ \\
        \hline
        MAE $\downarrow$
        & 9.215 & 9.576 & 9.953 & 10.30 & 10.12
        &
        & 15.92 & 16.29 & 16.69 & 17.12 & 16.77 \\
        MAE (ROI) $\downarrow$
        & 31.55 & 32.95 & 34.46 & 35.89 & 36.35
        &
        & 55.11 & 56.36 & 57.86 & 59.56 & 60.49 \\
        SSIM $\uparrow$
        & 0.9722 & 0.9704 & 0.9685 & 0.9666 & 0.9677
        &
        & 0.9429 & 0.9412 & 0.9393 & 0.9373 & 0.9394 \\
        SSIM (ROI) $\uparrow$
        & 0.8978 & 0.8908 & 0.8834 & 0.8761 & 0.8741
        &
        & 0.7957 & 0.7900 & 0.7833 & 0.7760 & 0.7728 \\
        \hline
    \end{tabular}
    \label{tab:angle_range}
\end{table*}

Finally, assuming clinical use of the proposed model, we conducted an evaluation experiment of CLCM using LACT images with different projection angle ranges simultaneously. Furthermore, by using LACT images with projection angle ranges not used during training for testing, we investigate generalization performance to imaging conditions anticipated in clinical settings.

As new training data, we prepared settings for guiding LACT images with the following LA acquisition and Extreme LA acquisition settings.
\begin{itemize}
 \item[a)] LA setting: $\theta_r = 120^\circ, 90^\circ, 60^\circ$
 \item[b)] Extreme LA setting: $\theta_r = 45^\circ, 30^\circ, 15^\circ$
\end{itemize}

We constructed a training database mixing all LACT images captured with three different projection angle ranges, and trained the model without specifying the angles. We applied LACT test images with each angle range to the trained model and evaluated the errors of the resulting sCT images. Furthermore, to evaluate the reconstruction performance for unseen angle ranges not included during training, we performed inference to obtain sCT images from LACT images with $\theta_r = 105^\circ$ and $75^\circ$ for the trained model of a), and from LACT images with $\theta_r = 37.5^\circ$ and $22.5^\circ$ for the trained model of b). This experimental setup enabled construction of more versatile models with expanded target projection angle ranges and verification of the reconstruction performance for LA settings that were different from the training conditions.

\begin{figure*}[t]
    \centering
    \includegraphics[width=0.95\linewidth]{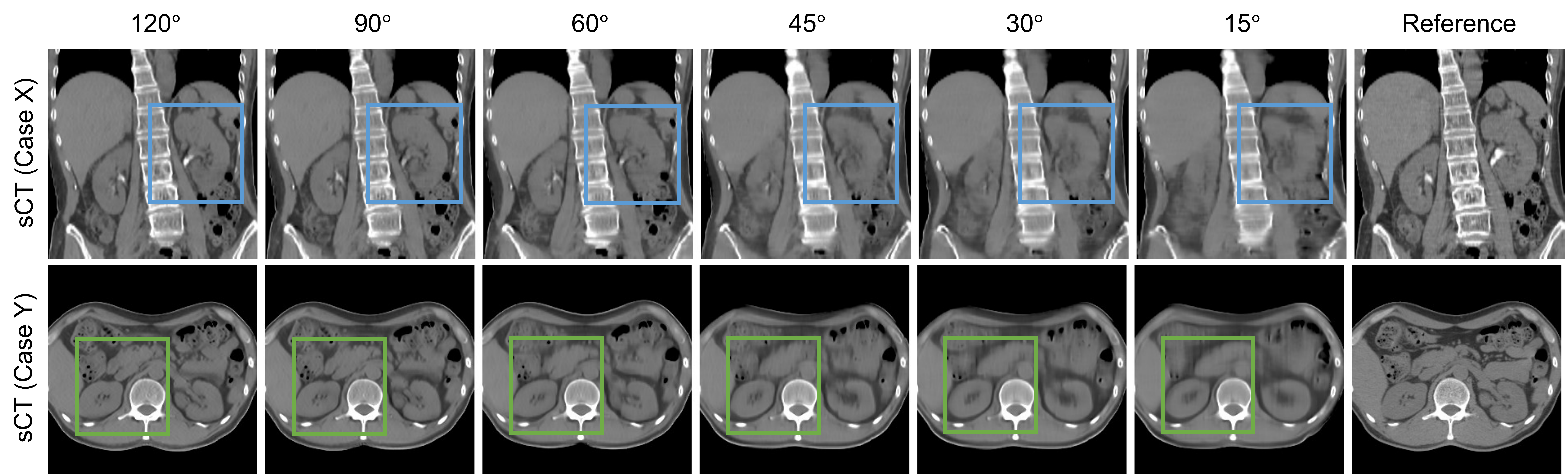}
    \caption{Synthetic CT images reconstructed from LACT inputs with various projection angle ranges. Reproducibility of major anatomical structures was maintained, and robust reconstruction is possible against changes in projection angle range.}
    \label{fig9}
\end{figure*}

Table \ref{tab:angle_range} shows the median evaluation metrics between sCT images obtained from ten types of LACT and reference images. Additionally, Figure \ref{fig8} plots the evaluation values of sCT images at each range of projection angle. Region (a) shows evaluation values of sCT images generated by the model trained with an LA setting, and region (b) shows those trained with an extreme LA setting. There is a notable difference in quality occurring at the boundary. In region (a), MAE increased with decreasing $\theta_r$, but the increase is suppressed compared with the loss of information included in the guidance. In particular, the MAE of sCT images in Experiment 2 obtained for LA images with $60^\circ$ was 10.29 [HU], whereas the model trained with the LA settings of $\theta_r = 120^\circ, 90^\circ$ and $60^\circ$ achieved 10.12 [HU], confirming an improvement in reconstruction error. This suppression of evaluation value deterioration was also confirmed in region (b), demonstrating the effectiveness of self-training restoration from LACT images with multiple projection angle ranges using a single model, rather than learning specialized for a specific angle range. 

Figure \ref{fig8} confirmed that the MAE increase maintains an approximately linear behavior, even for unseen projection angle ranges of $\theta_r = 105^\circ, 75^\circ, 37.5^\circ$, and $22.5^\circ$. We can thus confirm that the proposed model possesses generalization performance that achieves stable image generation, even for ranges of projection angles not included during training. 
Figure \ref{fig9} shows sCT images obtained in order of $\theta_r$ for three randomly selected cases from the test data. Reproducibility of major anatomical structures was maintained, which confirmed that robust reconstruction is possible against changes in projection angle range.

\section{Discussion}
This study proposed a multi-volume latent diffusion model that leverages latent representations derived from multiple field of views (FOVs) as guidance for LACT reconstruction. Through quantitative evaluation, we demonstrated that a model trained to learn the reconstruction task from LACT images with multiple projection angle ranges achieved higher performance than models specialized for a specific projection angle range. Furthermore, the proposed model exhibited stable image generation performance even for unseen projection angle ranges not included in the training data, indicating generalization capability. In this study, projection angle ranges from $15^\circ$ to $120^\circ$ were investigated, while the projection direction was fixed around the anterior-posterior orientation. Extension to arbitrary projection angle ranges and projection directions encountered in clinical settings, as well as further validation of reconstruction performance under such conditions, remain important areas for future investigation.

The experiments conducted in this study were limited to simulation-based analysis in which sCT images were generated from DRRs. Clinical evaluation of reconstruction performance using actual X-ray images remains an important future task. DRR and real X-ray images differ in terms of spatial resolution, field of view, and image characteristics. In addition, clinical imaging involves acquisition noise and device-dependent variations arising from differences in imaging systems. Investigating the impact of such domain discrepancies between training and inference phases, and their effects on the generated sCT images, will be essential for future clinical translation.

\section{Conclusion}
\label{sec:conclusion}
In this study, we proposed an image reconstruction model combining multi-volume encoding with a latent consistency model and its self-supervised learning method for high-precision image reconstruction under LACT conditions. Through a framework that encodes sub-volumes of the entire image and central region, we enabled efficient learning of high-frequency structures inside organs and 3D boundary information that were difficult to capture with conventional single-scale latent representations. Furthermore, by extending the consistency model to latent space, we achieved stable image generation in a single step.

We confirmed via evaluation experiments that the proposed model achieved sCT image generation with a quality surpassing other existing models. From self-training restoration from LACT images obtained with multiple projection angle ranges using a single model, we demonstrated that sCT image generation was possible with a MAE of 10.12 [HU] and a SSIM of 0.9677 under the LA acquisition setting of $60^\circ$, and with a MAE of 16.69 [HU] and a SSIM of 0.9393 under the extreme LA setting of $30^\circ$. We confirmed that the proposed model shows stable reconstruction performance, even for unseen LA settings not included during training, indicating its applicability to diverse clinical imaging conditions.

\section*{Acknowledgement}
We thank Edanz (https://jp.edanz.com/ac) for editing a draft of this manuscript.

\bibliographystyle{IEEEtran}
\bibliography{reference}

\end{document}